Research Article

# Do Socialization Restrictions Prevent Restaurants from Becoming Covid Hotspots?

**Aviral Bhatnagar\*, Himanshu Kharkwal and Jaideep Srivastava**

*Department of Computer Science, University of Minnesota, USA*

**\*Corresponding Author:** Aviral Bhatnagar, Department of Computer Science, University of Minnesota, USA.

## Abstract

Simulation models for infection spread can help understand what factors play a major role in infection spread. Health agencies like the Center for Disease Control (CDC) can accordingly mandate effective guidelines to curb the spread. We built an infection spread model to simulate disease propagation through airborne transmission to study the impact of restaurant operational policies on the Covid-19 infections. We use the Wells-Riley model to measure the expected value of new infections in a given time-frame in a particular location. For the purpose of this study, we have restricted our analysis to bars and restaurants in the Minneapolis-St. Paul region. Our model helps identify disease hotspots within the Twin Cities and proves that stay-at-home orders were effective during the recent lockdown, and the people typically followed the social distancing guidelines. To arrive at this conclusion, we performed significance testing by considering specific hypothetical scenarios. At the end of the study, we discuss the reasoning behind the hotspots, and make suggestions that could help avoid them.

**Keywords:** Simulation; Covid-19; Wells-Riley Model; Hotspot Detection

## Introduction

The Covid-19 pandemic has not only impacted our physical health[1] but has also disrupted mental health [1,7,12] and the global economy [2,3,6]. As many as 110,000 restaurants have closed since the pandemic was declared, food service sales have fallen $255 Billion this year[2], a contraction of this degree never seen before. Health agencies like the Center for Disease Control (CDC) and the Minnesota Department of Health (MDH) had mandated stay-at-home orders for most of 2020 and the first half of 2021 [4]. A pandemic of such a degree has not occurred for more than 100 years since the Spanish flu. As a result, there are no preceding accounts to learn from and accordingly take measures to curb the spread of the virus. Due to its novel nature, the CDC and other agencies have repeatedly amended their orders depending on the circumstances.

Extensive research has been carried out to model indoor infection spread. Covid-19 being an airborne infection, such studies provide an efficient method in simulating infection spread during the pandemic. We present a simulation toolkit that considers stochastic arrivals of people in a given period in restaurants and bars across the city. To evaluate the expected value of new infections, we use an infection transfer model that considers temporally and spatially co-located visits to a location where infected people can contribute to the spread. This model is called the Wells-Riley model. The Wells–Riley equation was developed by Riley and colleagues in an epidemiological study on a measles outbreak [8,9]. The equation is based on the concept of 'quantum of infection' [10]. We estimate the effectiveness of hypothetical infection control measures by modifying the parameters involved in our model through significance testing. For the purpose of this study, we restrict our analysis to the bars and restaurants within the Minneapolis-St. Paul region.

---

[1] http://mayoclinic.org/diseases-conditions/coronavirus

[2] https://www.restaurant-hospitality.com/operations/one-year-covid-19-pandemic-continues-impact-independent-restaurants

## Methods
### Dataset

We acquired data on the stochastic visits of people to bars and restaurants from the SafeGraph dataset[3]. The data set includes 1034 establishments that either come under the category "Restaurants and Other Eating Places" or "Drinking Places (Alcoholic Beverages)". We consider the time-frame starting from November 2nd, 2020, to November 9th, 2021, with visits being considered for every single hour of the week. SafeGraph data reports approximately 10% of the total volume of factual data. To address this sparsity, we multiply all the values in the data set by 10. We calculated the volumes by multiplying the area of each establishment with a constant height. The final output was then spatially joined with a different data set that kept a record of visits for each hour. We assume that the visitors do not spend more than an hour in the restaurant, and leave the premises before the new hour begins.

### Initial prevalence

The initial prevalence rate of infection within a given community represents the ratio between infected people and the total population, as shown in equation 1. In an infection spread model, this rate needs to be taken into account. The state of Minnesota witnessed a high prevalence rate between November 2nd and November 9th. However, the number of undocumented cases in Minnesota were approximately 15 times higher than the documented cases [11]. We incorporated these values into our model to calculate the proportion of initially infected people in every visit.

$$\text{Prevalence Rate} = \frac{\text{Number of Infected People}}{\text{Total Population}} \qquad 77 \quad (1)$$

### Indoor transmission

We have restricted our study to the indoor transmission of the virus through air. We employ the Wells-Riley equation to calculate the expected value of newly infected people within a given hour. It provides us with the probability of each susceptible individual getting infected when in the vicinity of an infected person. Several parameters determine the probability of the spread, such as the volume of the room, ventilation rate, time spent, pulmonary ventilation rate, and the initial prevalence. The Wells-Riley [8-10] equation is given below:

$$P = \frac{C}{S} = 1 - e^{\frac{-Iqpt}{Q}} \qquad (2)$$

P- Probability of infection spread

C- Number of newly infected people

S- Number of susceptible people

I- Number of infectors

p- Pulmonary ventilation rate of susceptible people

Q- Room ventilation rate

q- Quantum generation rate

t- Exposure time.

The pulmonary ventilation rate is set at 0.48 m$^3$/h [5]. The quantum generation rate is set at 20 quanta/h. The height of each establishment is assumed to be 3 meters. The initial prevalence of the infection is taken from covidactnow[4]. The room ventilation rate is the product of the ventilation rate 4 ac/h and the volume of the room.

## Results and Discussion

For the purpose of this study, we categorized infection spread into two categories: mild and severe. Severe incidents are defined as the establishments that witness more than one new infection in the week. On the other hand, mild incidents are the ones with less than one new infection that week.

---

[3]https://www.safegraph.com/

[4]covidactnow.org

With the stay-at-home orders in place, our model identified nine venues to have contributed to more than one new infection that week. However, the majority of new infections remained below 1, as shown in figure 1a. This figure is a histogram that shows a log-normal distribution of expected values of new infections. The second figure is a histogram representing severe cases. These values were calculated using equation 2.

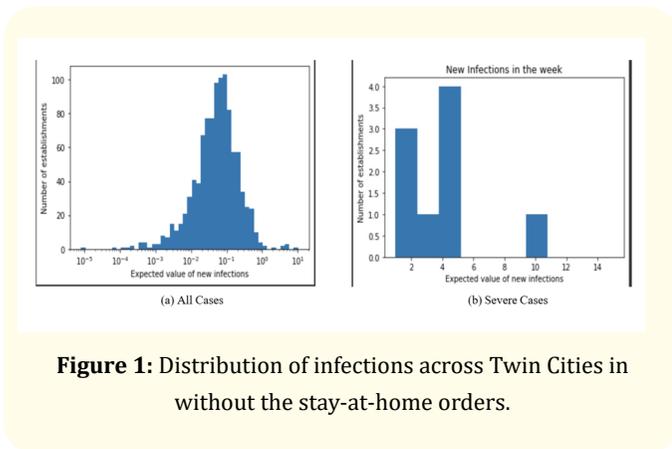

**Figure 1:** Distribution of infections across Twin Cities in without the stay-at-home orders.

### Hypothetical scenario 1: No stay-at-home orders in place

To simulate this scenario, we collected data on arrivals from 2019, when there were no stay-at-home orders in place. To ensure consistency, we considered the same time of the year and kept all the other parameters the same. As figure 2a and 2b suggest, we notice a significant shift of the graph to the right. The expected values are considerably greater, and several establishments had severe infection spread. To be precise, there were 116 venues with more

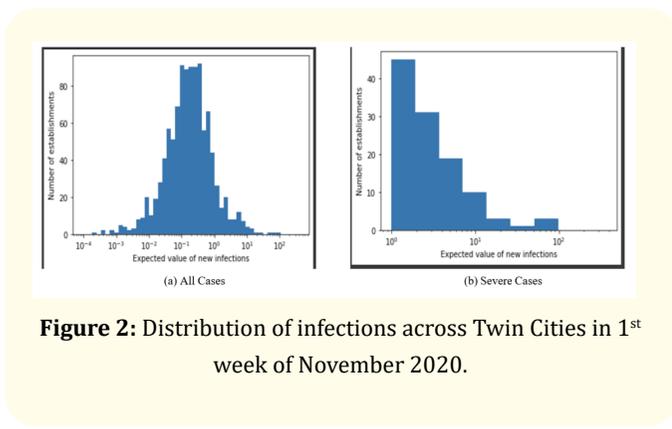

**Figure 2:** Distribution of infections across Twin Cities in 1st week of November 2020.

### Significance testing

Superimposing both the graphs together, as shown in figure 3, we notice a considerable amount of right shift when the stay-at-home orders are lifted. To quantify this difference, we perform a significance test to check whether the stay-at-home orders significantly reduced the infection spread.

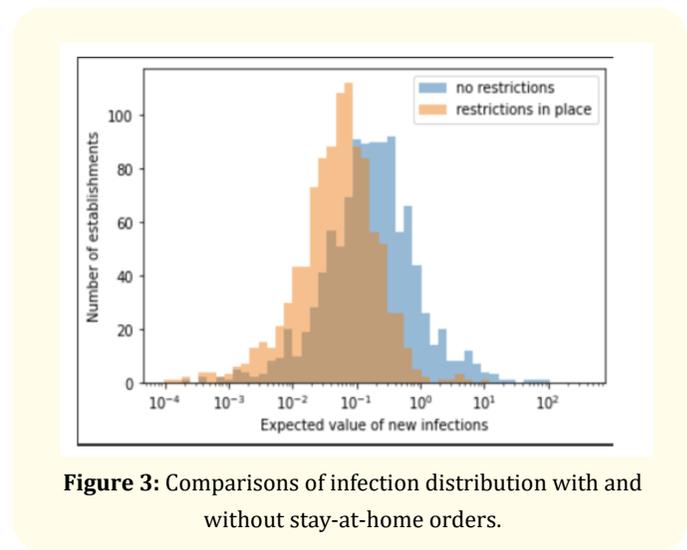

**Figure 3:** Comparisons of infection distribution with and without stay-at-home orders.

After performing the t-test, we get a p value of 0.006, which means that there is a 0.6% chance that a difference in the infection rates is as extreme as the two distributions shown above due to random chance. Considering a 99% confidence level, we can conclude that the difference is significant, and the stay-at-home orders were effective in curbing infection spread.

### Hypothetical scenario 2: Physical distancing

The CDC had advised people to stay 6 feet apart in public places[5]. When a person is in close contact (within 6 feet) for a prolonged period, the infection can spread when the infected person coughs, sneezes, or talks, and the droplets from their mouth or nose are launched into the air[6].

---

[5]cdc.gov/coronavirus/2019-ncov/prevent-getting-sick/social-distancing.html

[6]https://www.cdc.gov/coronavirus/2019-ncov/prevent-getting-sick/social-distancing.html

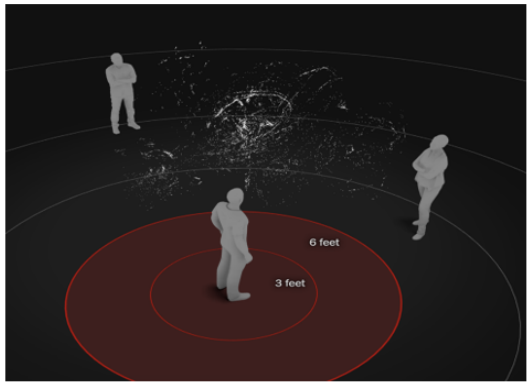

**Figure 4:** Visual representation of 6 feet social distancing.

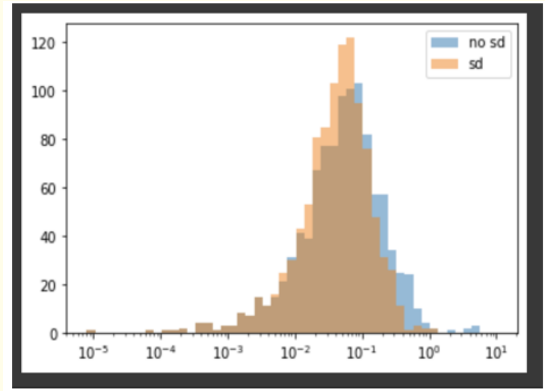

**Figure 5:** Comparison of Infection distribution between ideal and actual scenarios.

Our simulation creates a hypothetical environment where the 6 feet social distancing rule is strictly enforced. The flexibility of our framework enables us to simulate such an environment. It would prevent establishments from over saturating the occupancy by following the guide-lines. For example, if we consider a hypothetical restaurant that is circular with a radius of 6 feet, such a restaurant would permit only one person at any given hour, as shown in figure 4. Our framework would permit new visitors until the maximum permitted occupancy is reached.

As seen in figure 5, there is a subtle right shift on the x-axis, implying that if the social distancing rule were strictly enforced, the expected value of infection spread might have been lower. The significance test, however, gave an extremely large *p value* of 2.42. This concludes that the difference in the new infection rates is not significant and indicates that the residents of Minneapolis-St. Paul typically abode by the social distancing rule.

### Future work

Since the onset of Covid-19 it has caused unimaginable disruption in the society. The guidelines provided by administrators helped us to navigate through the pandemic. We want to create a simulation framework for policy makers to implement and study the impact of their operational policies. In this study we analyzed the impact of policies such as stay at home orders and physical distancing. In future we plan to implement policies such as limited hours of operations, maximum occupancy, limited group size and the duration they spend inside the restaurant, air circulation, and frequency of cleaning. Further, we want to validate our results against the ground truth of infections.

The policy decisions help in controlling the infection spread but have a socio-economic cost associated with them. Covid-19 has somewhat measurable impact on health, but it is difficult to quantify the socio-economic cost. The decisions such as quarantining hotspots will significantly reduce the infection spread but will create an economic burden. We want to study the economic impacts of such decisions in future.

### Conclusion

We leveraged the flexibility of the Wells-Riley equation which enables us to simulate a hypothetical environment, proving to be helpful to test new or proposed guidelines for similar circumstances in the future. This study was conducted using real data from the Minneapolis-St. Paul region for the first week of November 2020. The impact of changing guide-lines was assessed by simulating hypothetical environments and testing them against the real data. We used the results to get a measure of confidence in each decision. We concluded that the lockdown measures were effective in curbing the spread of the disease, and people in the Twin Cities generally followed the social distancing guidelines. However, in spite of strict social distancing measures, our model identified potential hotspots. These hotspots witnessed a large amount of traffic and enforcing strict social distancing measures was not sufficient. We suggest that these establishments could enforce stricter guidelines and conduct stringent policing to ensure that the guidelines are appropriately followed. Health agencies and other organizations can use this framework to assess the effectiveness of their policies. Moreover, even though the stay at home orders in conjunction with strict physical distancing work best to curb the infection spread, we speculate that such a strict mandate would cause severe socio-economic impact on local businesses. Instead, if we consider enforcing only the stay-at-home orders, they would still significantly reduce the number of new infection hotspots, while having a softer impact on the local economy, as shown in figure 3. Enforcing strict social distancing does not significantly reduce the number of infection hotspots further, as shown in figure 5.